\documentclass[twocolumn]{aastex63}
\usepackage{apjfonts}
\usepackage{amssymb, amsmath}
\usepackage{aas_macros, natbib, ulem}
\newcommand{\msun}{\rm M_\odot}

\newcommand{\mbulge}{M_{\rm bulge}}
\newcommand{\eg}{e.\,g.~}

\newcommand*\diff{\mathop{}\!\mathrm{d}}

\newcommand{\num}[2]{#1\times 10^{#2}} %scientific notation

\begin{document}
\title{Prospects for Memory Detection with Low-Frequency Gravitational Wave Detectors}
\author{K.~Islo}
\affiliation{Center for Gravitation, Cosmology and Astrophysics, Department of Physics, University of Wisconsin-Milwaukee,\\ P.O.~Box 413, Milwaukee, WI 53201, USA}
\author{J.~Simon}
\affiliation{Jet Propulsion Laboratory, California Institute of Technology, 4800 Oak Grove Drive, Pasadena, CA 91109, USA}
\author{S.~Burke-Spolaor}
\affiliation{Department of Physics and Astronomy, West Virginia University, P.O.~Box 6315, Morgantown, WV 26506, USA}
\affiliation{Center for Gravitational Waves and Cosmology, West Virginia University, Chestnut Ridge Research Building, Morgantown, WV 26505, USA}
\author{X.~Siemens}
\affiliation{Center for Gravitation, Cosmology and Astrophysics, Department of Physics, University of Wisconsin-Milwaukee,\\ P.O.~Box 413, Milwaukee, WI 53201, USA}
\correspondingauthor{K.~Islo}
\email{kpislo@uwm.edu}

\begin{abstract}
Gravitational wave memory is theorized to arise from the integrated history of gravitational wave emission, and manifests as a spacetime deformation in the wake of a propagating gravitational wave. We explore the detectability of the memory signals from a population of coalescencing supermassive black hole binaries with pulsar timing arrays and the Laser Interferometer Space Antenna (LISA). We find that current pulsar timing arrays have poor prospects, but it is likely that between 1 and 10 memory events with signal-to-noise ratio in excess of 5 will occur within LISA's planned 4-year mission.
\\
\end{abstract}

\section{Introduction}
\label{sec:intro}
Non-linearities in Einstein's field equations suggest that during the coalescence of compact objects, the oscillatory gravitational wave (GW) radiation is accompanied by a monotonically increasing or decreasing component of the strain. The result is an enduring change of gravitational potential in the wake of a propagating GW, termed ``memory'' \citep{Christodoulou91, Blanchet92, Thorne92}. For example, a GW with memory passing through a system of two isolated, free-falling test masses would permanently stretch or compress the comoving distance between them. 

Like the oscillatory component of a GW, memory is sourced by a changing time derivative of the system's mass multipoles, but only in part. It also grows through the cumulative history of GW emission. As such, the memory signal inherits the radiating system's evolving past: its strength at any time is the result of the integrated history of the system. For a supermassive black hole binary (SMBHB) undergoing coalescence, the memory signal initially displays negligible growth corresponding to the slow time evolution of the binary's inspiral. Once the binary enters its most dynamic phase during coalescence, the system emits a burst of memory signal which propagates outwards. Observations of SMBHB coalescence memory events would shed light on strong-field effects of General Relativity (GR), and provide information about SMBHB properties augmenting that obtained from the oscillatory components. More broadly, a memory observation would provide robust evidence of fundamental symmetries in GR \citep{Strominger}.

In this paper, we estimate the current and future potential to detect GW memory from SMBHB coalescence based on a simulation-suite of semi-analytic models for the SMBHB population. Importantly, the models are based in local observables for SMBHBs, and thus encompass only uncertainties from local mass functions, galaxy merger timescales, etc.\ rather than uncertainties in casting dark matter halo simulations to a binary supermassive black hole population (as has dominated such simulations in the past)\citep{Enoki2004, Sesana2008, Ravi2012, Kulier2015}. This study is also novel from previous studies in that we expand our models to include lower black hole masses, down to $M_{\rm BH} \gtrsim 10^5$, and higher redshifts, such that the memory signals expected in the bands relevant to both Pulsar Timing Arrays (PTAs) and the Laser Interferometer Space Antenna (LISA) can be explored. We also take into account the unknown decoupling radius for binary-host interactions. We look at the likely rate and signal-to-noise ratio (SNR) for memory events detectable within these frequency bands. 

This paper is structured as follows: Section~\ref{sec:SMBHB} contains motivation and description of our chosen source population, Section~\ref{sec:model_new} describes the GW memory signal model, Section~\ref{sec:results} reports the average memory burst rate for memory events of varying strength, and we summarize results in Section~\ref{sec:conclusions}.

\section{Supermassive Black Hole Binaries}
\label{sec:SMBHB}
The largest black holes in the Universe are created by hierarchical evolutionary processes involving the mergers of increasingly massive galaxies. SMBHBs form during major galaxy mergers, and grow more tightly bound through repeated interactions with their galactic environment, which drives orbital evolution to smaller separations \citep{Begelman+1980, Volonteri+2003}. The effectiveness of the mechanisms by which these black hole systems are driven to coalescence is an open question in astrophysics, although many studies conclude these binaries will coalesce given sufficiently gas-rich galactic environments or continuous rates of stellar loss-cone refilling \citep{Colpi2014, Roedig+2014, Khan+2013, Vasiliev+2015}. SMBHBs that eventually coalesce are candidates for producing strong GW memory bursts. 

The persisting spacetime offset induced by a passing GW memory signal is proportional to $\Delta E_{\rm rad}/D$, where $\Delta E_{\rm rad}$ is the total energy radiated in the form of GWs and $D$ is the comoving distance to the coalescence. For equal-mass SMBHBs, the energy available for the GW memory burst ranges from 5\% to 10\% of the total binary energy. The precise value depends on binary inclination and the degree of black hole spin-alignment (see \cite{pollney2010}). For example, an optimally-oriented binary consisting of two $10^{9} \, \msun$ black holes coalescing $1 \,\rm Gpc$ away from Earth will emit a GW memory burst with amplitude $h_{\rm mem} \sim 10^{-15}$.

\noindent
{\it Population synthesis:}
To find SMBHB coalescence rates we use the population synthesis model described in \citet{simonBS}. The number density of SMBHB coalescences occurring in a time interval $dt$ from galaxy pairs with redshift, mass, and mass ratio range $\diff z$, $\diff M$, and $\diff q$, respectively is

\begin{equation}\label{eq:MergerRate}
    \dfrac{d^4 N}{dt\,dM_{*}\,dq\,dz_{\rm burst}} = \left.\dfrac{d^3 n}{dz_{\rm gal}\,dM_{*}\,dq}\right|_{\rm gal \rightarrow \bullet}\, \dfrac{dV_{c}}{dz_{\rm burst}}\,\dfrac{d z_{\rm burst}}{dt}.
\end{equation}

The integrated value $N$ is parameterized by astrophysical, observable quantities that characterize galaxy mergers -- namely, galaxy stellar mass function (GSMF), galaxy pair fraction, and galaxy merger timescale. These are cast from galaxy pairs into inferred SMBHBs using the empirical relationship found between host galaxy bulge mass and black hole mass \citep{McConnellMa13, Shankar+2016}. The last factor in Eq.~\eqref{eq:MergerRate} is a cosmological term converting the proper time rate to a redshift rate. Assuming coalescence does not immediately follow binary formation, we must consider consider two redshifts: the redshift at which a galaxy merger forms a hardened binary $z_{\rm gal}$ and the redshift of the memory burst upon SMBHB coalescence $z_{\rm burst}$. 

We restrict the parameter space to include only what we know observationally. 
Specifically, we take the redshift to be $z < 3$, the primary galaxy mass to be $10^8 \,\msun \leq M_{*} \leq 10^{12} \,\msun$, and the mass ratio to be consistent with `major mergers' with $0.25 < q < 1$.

\noindent
{\it Evolution of SMBHBs:}
At the early stages of galaxy merger, we expect dynamical friction to reduce the orbital angular momentum of the individual black holes until they sink to the center of the merger remnant, forming a SMBHB \citep{Begelman+1980, Volonteri+2003}. The dominant mode of energy loss below $\sim 10 \,\rm pc$ binary separation is not yet understood, although many environmental interactions potentially contribute \citep[]{merritt-living-reviews-article}. As such, it is unclear when the environment decouples from the binary, after which GW emission dominates. This has been the center of a growing debate in predicting the strength and spectral shape of the stochastic gravitational wave background in the nanohertz frequency band \citep[\eg][]{nanograv_isotropic_11yr, Taylor+2017, Rasskazov2017, Kelley+2017}. 

Here, we incorporate varying environmental influence to predict GW memory bursts for populations with different evolutionary timescales. This will dictate the offset between $z_{\rm gal}$ and $z_{\rm burst}$.
We introduce a parameter we term the {\it decoupling radius} $a_d$ to be the orbital separation at which the binary's evolution typically becomes GW-dominated.
SMBHBs embedded within sparse environments exhaust their environmental interactions earlier and have the potential to stall before reaching a regime where GW-radiation can drive the binary to coalesce (i.e., $a_d \gtrsim 1 \rm \,pc$). The opposite scenario involves a binary strongly coupled to its environment, undergoing extremely efficient orbital shrinking, and reaching coalescence quickly ($a_d\to 0$).

If we assume circular binaries, the time to coalescence can be expressed in terms of Keplerian parameters and chirp mass $\mathcal{M} = (M_{1}M_{2})^{3/5}/(M_{1}+M_{2})^{1/5}$   
\begin{equation}
    \tau_{\rm GW} = \dfrac{5\,c^5}{256\,G}\, \dfrac{a_{d}^4}{\mathcal{M}^{5/3} M_{\rm tot}^{4/3}}\label{eq:tau_GW},
\end{equation}

Establishing $a_d$ is therefore akin to specifying the total time to binary coalescence, i.e., $t_{\rm burst} = t_{\rm gal} + \tau_{\rm GW}$, where $t_{\rm gal}$ is the time between galaxy merger and SMBHB formation.

We adopt a simple power-law model relating total binary mass and decoupling radius to emulate any environmental interaction that is more common among smaller SMBHBs than larger ones:

\begin{equation}
    a_{d} = a_{8}\left(\dfrac{M_{\rm tot}}{10^8\,\msun}\right)^{\alpha}    
\end{equation}

\noindent where $a_{8}$ is the decoupling radius corresponding to an SMBHB with total mass $10^{8}\, \msun$. 

The least efficient environments consist of a persistently depleted loss-cone and sparse gas in the galaxy merger core. Following \citeauthor{Begelman+1980}, we find the orbital separation at which the binary will stall for a given galaxy-merger-bulge mass. We assume the ratio between bulge radius and bulge mass to be linear, with M87 serving as the fiducial ratio. In this scenario, we find best-fit parameters  $a_{8}= 1.3 \, \rm pc$ and $\alpha=1.0$ to describe this minimally-efficient binary environment.

A maximally-efficient environment allows for even the most massive binaries to reach sub-parsec separations through continual loss-cone refilling and a ready supply of in-flowing gas. In this context, we choose $a_{8}=0.01\, \rm pc$ (per Figure 1 of \citeauthor{Begelman+1980}) and consider $1.0 \leq \alpha \leq 3.0$.

\noindent
{\it Addition of lower-mass black holes:} 
Previous simulations of SMBHB populations (such as that in \citeauthor{simonBS}) only include progenitor galaxy mass greater than $10^{10} \,\msun$ which corresponds to individual black hole mass of $\sim 10^{7} \,\msun$. Lower galaxy masses are likely not relevant PTA-band sources, but $10^{5} - 10^{7} \,\msun$ black hole binaries can produce GW memory signals which enter the LISA band. We use results from the Galaxy And Mass Assembly survey as outlined in \cite{Baldry+2017} to estimate the distribution of these lower mass binaries at $z < 0.1$ along with those from the ULTRAVista survey reported in \cite{Ilbert+2013} for higher redshifts.

\section{GW Memory Signal Model}
\label{sec:model_new}
 Over the binary's lifetime memory undergoes a slow growth prior to merger, rapid accumulation of power during coalescence, and eventual saturation to a constant value at ringdown. Thus, in the time domain, the signature of a memory signal from a SMBHB can be approximated by a step-function centered at the moment of coalescence, i.e., 
 
\begin{equation}\label{eq:step}
	h_{+,\times}^{(\textrm{mem})}(t) = \Delta h_{+,\times}^{(\textrm{mem})} \Theta(t), 
\end{equation}

\noindent where $\Theta$ is the Heaviside-step function. In the frequency domain,

\begin{equation}\label{eq:h_tilde}
\tilde{h}_{+,\times}^{(\textrm{mem})}(f) = 
\begin{cases}
\dfrac{i \Delta h_{+,\times}^{(\textrm{mem})}}{2 \pi f} & \text{for $0 < f < f_{c}$}, \\
0 &\text{for $ f \geq f_{c}$},
\end{cases}
\end{equation}
where $f_{c}$, the \textit{cut-off frequency}, corresponds to twice the orbital frequency at coalescence. Frequencies larger than $f_{c}$ do not contribute to the GW signal. We can approximate $f_{c} \sim {\tau}^{-1}$, where $\tau$ is the light crossing time of the merger remnant. $\tau$ is also the timescale for the rise of the memory signal during the merger.
\eqref{eq:h_tilde} is sufficient for PTA signal-to-noise ratio calculations, however, we include a minor correction since LISA may be able to resolve the time varying features of the memory signal between onset of coalescence and ringdown. From \cite{Favata}:
\begin{align}
	\tilde{h}_{+}^{(\textrm{mem})}(f) \approx i \dfrac{\Delta h^{(\textrm{mem})}_{+}}{2 \pi f}\left[1 - \dfrac{\pi^2}{6}(\tau f)^2\right]\label{eq:hmem_f}.
\end{align}
The magnitude of the spacetime offset $\Delta h_{\rm mem}$ is affected by black hole spin-alignment, with the maximally aligned spinning case exhibiting the strongest signal \citep{pollney2010}. Higher-order spin effects should be incorporated in future simulations to properly reflect the saturated memory amplitude. We use the spin-averaged formula as given by \cite{Madison17}, 
\begin{equation}\label{eq:h_plus}
    \Delta h_{+}^{(\textrm{mem})} \simeq \dfrac{(1 - \sqrt{8}/ 3)}{24}\, \dfrac{G\mu}{c^2 D} \sin^2 \left(17+ \cos^2 \iota\right),
\end{equation}
with dependence on additional binary parameters reduced mass $\mu$, inclination angle $\iota$, and comoving distance $D$.

\begin{figure}[htb]
\centering
        {%
            \includegraphics[width=\columnwidth]{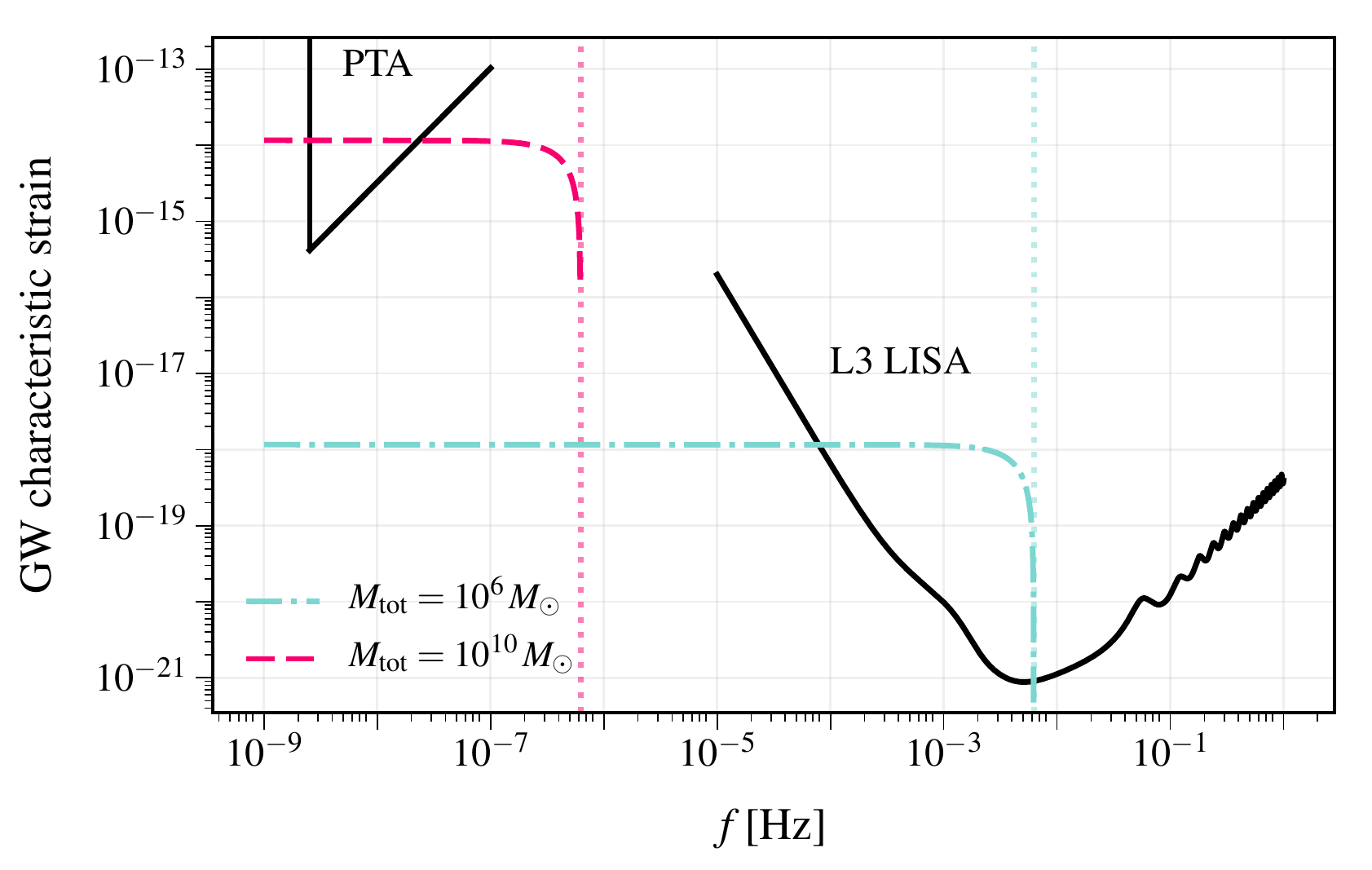}
        }%    

    \caption{LISA noise curve as of the L3 proposal \citep{Amaro-Seoane2017} (orange) along with an example PTA with a baseline of 12.5 years and rms residual errors of 100 ns (blue). We use Eq.\, \ref{eq:hmem_f} and Eq.\,\ref{eq:h_plus} to plot the GW memory signals from two equal-mass SMBHBs with total mass $10^{10} \,\msun$ and $10^{6} \,\msun$, at $z=1.0$ in green and red, respectively.}
\label{fig:joint_curve}
\end{figure}

\subsection{Signal-to-Noise Ratio}
Previous studies expound on ideal strategies for searching and conducting analyses of memory signals (e.g. \cite{Braginsky+Thorne}). In this work, rather than discuss best practices, we evaluate the astrophysical motivation for implementing searches of GW memory signals from a cosmic population of SMBHBs, specifically focusing on the signal-to-noise for memory events within LISA and PTAs.

\subsubsection{LISA}
Proposed space-based interferometers like LISA would respond to a burst with memory as a permanent change to the proper distance between its free mirrors. In theory, the signal may be stored forever. Ground-based interferometers are not optimized for memory detection, requiring stacking of sub-threshold stellar-mass coalescence signals \citep{Lasky2016}, or waiting on new space-based observatories like the Big Bang Observatory to detect memory from GW150914-like events \citep{Johnson+2019}. Here we calculate the SNR for memory from higher mass binaries, relevant for LISA.

\begin{figure}[htb]
\centering
        {%
            \includegraphics[width=\columnwidth]{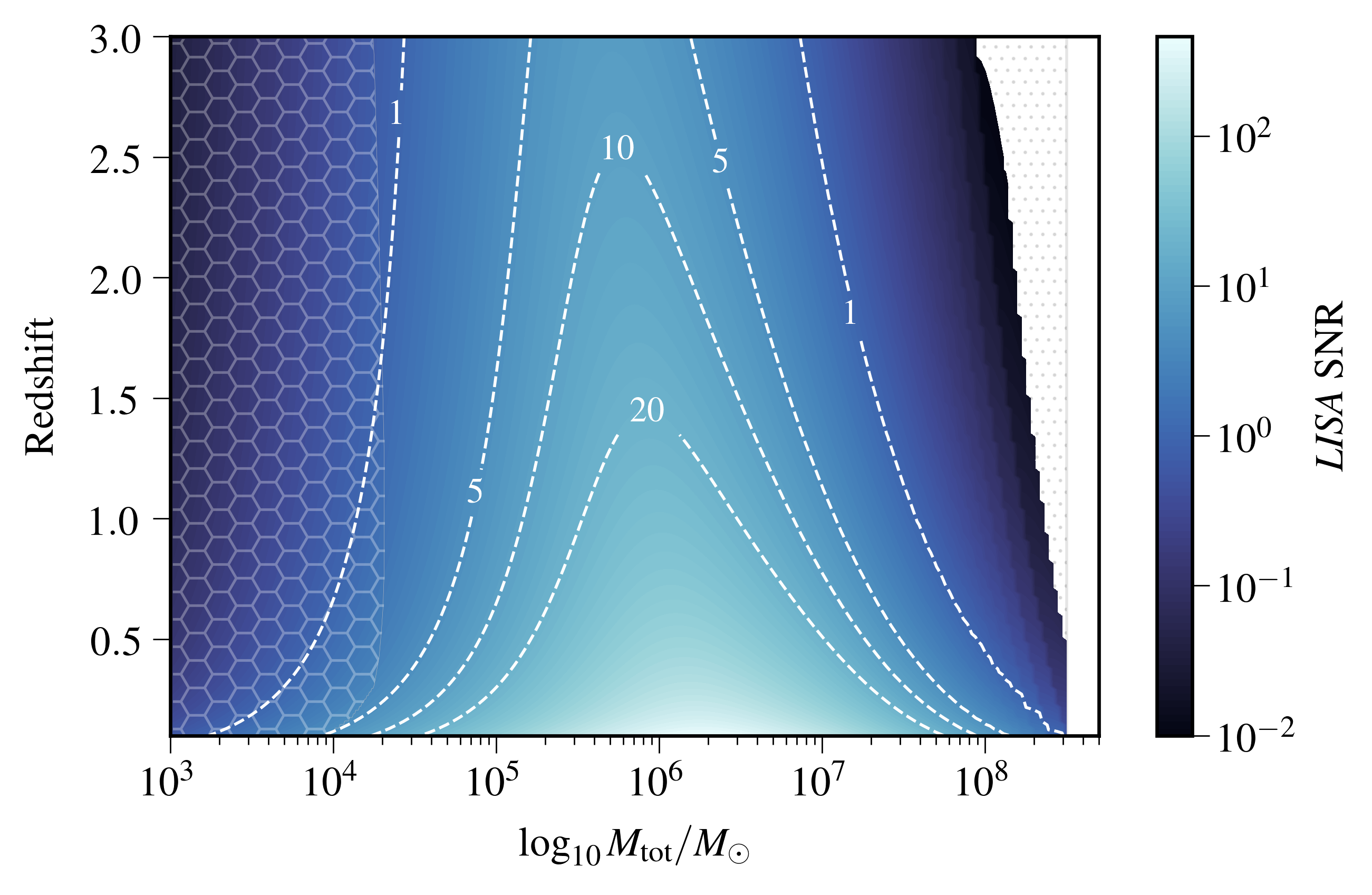}
        }%    
    \caption{LISA signal-to-noise ratio for GW memory bursts from optimally-beamed (e.i., $\iota = \pm \pi/2$), equal-mass MBHBs with $M_{\textrm{tot}}$ between $10^{3}\,\msun -  10^{9}\,\msun$ and coalescence redshift between $0.1 - 3.0$. Dots fill the space in which the memory burst signal falls outside the LISA band due to strain amplitude below the noise amplitude or a cut-off frequency below the minimum band frequency. The honeycombed region encompasses binaries for which the memory signal will be observable, while the oscillatory component of the signal at coalescence will not. Oscillatory merger frequencies determined via \citet{Robson+2019}.}
\label{fig:SNR_contour}
\end{figure}

The signal-to-noise ratio (SNR) is computed using an optimal match filter to extract a known signal model from noisy data as described in \cite{hughes1998}. With \eqref{eq:hmem_f} as our memory signal model, the average SNR is given by

\begin{align}
	\langle \rho^2 \rangle &= 4 \operatorname{Re} \int_0^\infty \dfrac{\langle \tilde{h_{+}}^{(\rm mem)}(f) | \tilde{h}_{+,(\rm mem)}^*(f) \rangle}{S_{f}} df, \\	
&= \dfrac{\left(\Delta h_{+}^{(\textrm{mem})}\right)^2}{\pi^2}\int_{0}^{f_{c}}\dfrac{\!\mathrm{d}f}{f^2}\left(1 - \dfrac{\pi^2}{6}(\tau f)^2\right)^2 \dfrac{1}{S_{n}(f)}. \label{eq:SNR_LISA}
\end{align}
Figure~\ref{fig:SNR_contour} shows the SNR for a memory burst produced by SMBHBs of total binary mass $M_{\rm tot}$ coalescing at redshift $z$. Expected SNR ranges from 100 to 10,000 with the highest SNR events coming from binaries at $z < 0.5$ and with $ 10^{5} \,\msun< M_{\rm tot} < 10^7\,\msun$. Our approximated memory model only minimally diverges at higher redshift from the LISA SNRs reported in \cite{Favata}. Analysis of LISA mergers within the same range of mass and redshift, indicate that $M_{\rm tot} < 10^{4.2}\,\msun$ coalescences will occur beyond the LISA frequency band ($\geq 1$ Hz), in which case, the memory signal will be the dominant \textit{coalescence} signature. ``Orphan memory'' signals, as they are termed in \cite{McNeill+2017}, effectively increase the high-frequency limit of the detector, allowing access to previously forbidden binary masses and redshifts.

\subsubsection{Pulsar Timing Arrays}
PTAs measure the times of arrival of radio pulses from an array of galactic millisecond pulsars. GW memory manifests in pulsar timing observations as an abrupt increase or decrease in pulsar rotational frequency. A burst passing Earth will affect the timing residuals for all pulsars in the array in a correlated way. Leveraging the data from many pulsars is crucial when claiming a confident memory detection within PTA data as intrinsic noise in individual pulsars can mimic a burst. Below we derive the SNR of a memory event within a single pulsar's residuals -- the SNR using an entire array is easy to estimate from this single pulsar derivation. 

If a burst signal like that expressed in \eqref{eq:step} spontaneously alters the $j$th pulsar's measured rotational frequency at time $t_{\rm mem}$, the Fourier transform of the integrated time-series residuals is

\begin{equation}\label{eq:residual_FF}
\tilde{R}_{j}(f) = - B_{j} \Delta h_{+}^{(\rm mem)} \dfrac{e^{ift_{\rm mem}}}{2 \pi f^2},
\end{equation}
where $B_{j}$ is a geometrical factor accounting for pulsar position, GW polarization and propagation direction (see \cite{VanHaasteren} or \cite{Madison+2014}).
PTA SNR is similarly determined through a matched filter analysis. From \cite{VanHaasteren}:

\begin{equation}\label{eq:SNR_PTA_0}
\langle \rho_{j}^2 \rangle = 4 \operatorname{Re} \int_{1/T_{\rm obs}}^{\infty} \dfrac{|\langle \tilde{R}_{j}(f) | \tilde{R}_{j}^*(f) \rangle|}{S_{r}(f)} \mathop{}\!\mathrm{d}f,
\end{equation}
where $S_{r}$ is the power-spectral density of the timing residuals and $T_{\rm obs}$ is the total time span of pulsar observation. Assuming the residuals are white and Gaussian, the one-sided power-spectral density can be written as

\begin{equation}\label{eq:S_white}
    S_{r}(f) = 2 \sigma^2 \Delta t, 
\end{equation}
where $\sigma$ is the white noise root-mean-square errors on the pulsar residuals and $1/\Delta t$ is the observation cadence (typically $\sim1$/month). Inserting \eqref{eq:residual_FF} and \eqref{eq:S_white} into \eqref{eq:SNR_PTA_0}, and averaging over pulsar and source location gives

\begin{equation}\label{eq:SNR_PTA_1}
\textrm{SNR} = \dfrac{\Delta h^{(\rm mem)}_{+} T_{\rm obs}^{3/2} }{6\, \pi \sigma \Delta t}.
\end{equation}

If a GW memory wavefront were to strike the Earth, we can correlate the entire array's timing residuals over the burst event epoch to uncover frequency changes in every pulsar. Multiple pulsar observations enables more confident GW memory detection by increasing the observed SNR by a factor proportional to $\sqrt{N_{\rm pulsars}}$ (\citeauthor{VanHaasteren}). 
We incorporate \eqref{eq:SNR_LISA} and \eqref{eq:SNR_PTA_1} into the simulation outlined in \ref{sec:SMBHB} to repeatedly generate instances of an ensemble of coalescing SMBHBs, allowing us to arrive at GW memory event statistics for PTAs and LISA. 

\section{Results}
\label{sec:results}
For clarity, we provide a glossary describing the characteristics of the 4 extremal cases of binary parameters we report on below (see Table \ref{tab:models}). These include minimal and maximal stalling and the inclusion or exclusion of potential selection bias when constructing an $M_{\rm BH}-M_{\rm bulge}$ relation. 

\begin{center}
\begin{table*}[htb]
    \scriptsize
    \caption{Glossary of decoupling models}
    \begin{center}

    \begin{tabular}{@{}cccccccc@{}}
    \hline\hline
       \multicolumn{1}{c}{Model} & \multicolumn{1}{c}{GSMF} && \multicolumn{1}{c}{$M-\mbulge$}
        &&\multicolumn{1}{c}{Orbital Decay}
        &&\multicolumn{1}{c}{Power-law parameters}\\
            \hline
            A           & Ilbert+Baldry && McConnell \& Ma &&
            replenished loss-cone; gas-driven && $1.0\leq\alpha\leq3.0$,  $a_{8} =0.01\,\rm pc$\\
            B          & Ilbert+Baldry && Shankar  &&
            replenished loss-cone; gas-driven && $1.0\leq\alpha\leq3.0$,  $a_{8} =0.01\,\rm pc$\\
            C           & Ilbert+Baldry && McConnell \& Ma && no loss-cone refilling, no gas && $\alpha=1.0$,  $a_{8} =1.3\,\rm pc$\\
            D           & Ilbert+Baldry && Shankar && no loss-cone refilling, no gas && $\alpha=1.0$,  $a_{8} =1.3\,\rm pc$\\
            \hline
    \end{tabular}
    \end{center}

    \label{tab:models}
\end{table*}
\end{center}
\begin{figure}[htb]
\centering
        {%
            \includegraphics[width=\columnwidth]{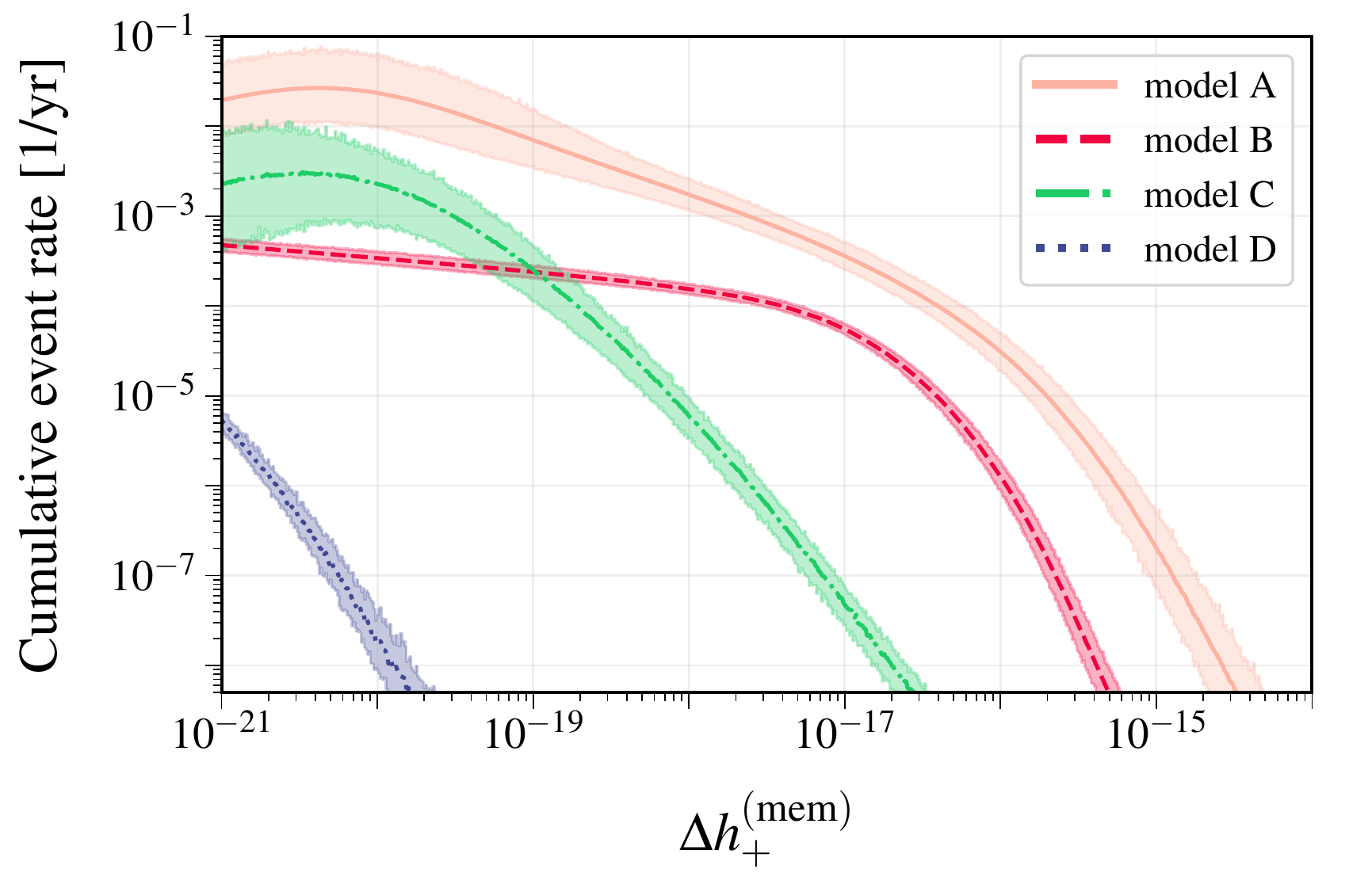}
        }%    

    \caption{Cumulative memory burst rate for bursts with strain amplitude at or above strain amplitude $\Delta h^{(\rm mem)}_{+}$ as in \eqref{eq:h_plus} for 100 realizations of SMBHB populations. Bold lines in models A and B indicate the mean across simulations using $1.0 \leq \alpha \leq 3.0$; the shaded regions encapsulates $1-\sigma$ range across these sub-models. Models C and D show a median rate and $1-\sigma$ interval.}
\label{fig:rate_vs_strain}
\end{figure}

\begin{figure*}[ht]
\centering
        {%
            \includegraphics[width=\textwidth]{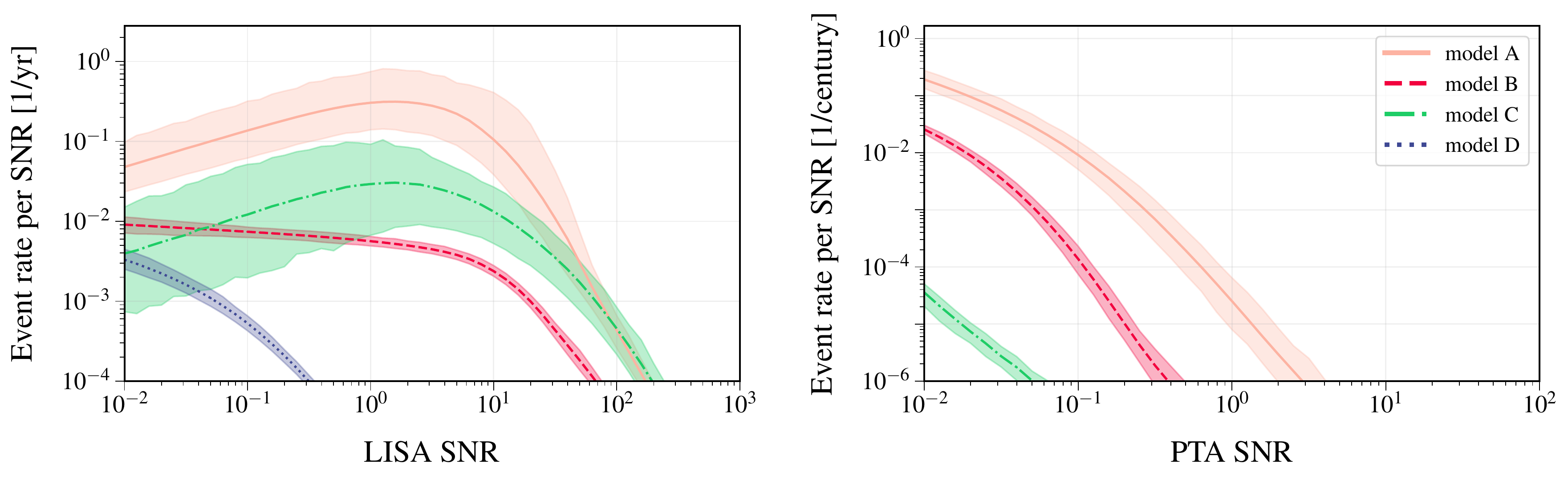}
        }%    

    \caption{Mean number of memory bursts per SNR bin from 100 realizations of SMBHB populations assuming outlined models. Shaded regions encapsulate $1-\sigma$ intervals. Bold lines in models A and B indicate the mean across median rates using $1.0 \leq \alpha \leq 3.0$; the shaded regions encapsulates $1-\sigma$ range across these sub-models. Models C and D show a median rate and $1-\sigma$ interval. NB: The axes' units.}
\label{fig:rate_vs_snr}
\end{figure*}

Figure~\ref{fig:rate_vs_strain} shows $\Delta h^{(\rm mem)}_{+} \geq 5 \times 10^{-16}$ can occur between $\num{8.7}{-4}$ and $\num{5.2}{-3}$ times per century for model A and $\num{2.9}{-6}-\num{1.4}{-5}$ times per century for model B. Models C and D approach rates of less than once per Hubble time for these stronger memory bursts. For assembled SMBHB populations initiated from $10^{10}\,\msun \leq M_{*} \leq 10^{12}\,\msun$ and z < 3, \citet{Ravi+15} finds such an event to occur once per century. We recover the results of \citeauthor{Ravi+15} if we assume PTA-relevant binaries have decoupling radii less than $0.001 \,\rm pc$. Evidence suggests this assumption may be too optimistic as estimates from \textsc{Romulus25} point towards a relatively large fraction of major-merger SMBHBs formed by $M_{\rm Gal} > 10^{9} \msun$ requiring Gyr timescales to evolve to sub-parsec separations \citep{Tremmel2018}. 

We use \eqref{eq:SNR_LISA} and \eqref{eq:SNR_PTA_1} to calculate the SNRs associated with each memory burst in Figure~\ref{fig:rate_vs_strain}.
Figure~\ref{fig:rate_vs_snr} shows the mean number of burst events per year of given SNR specific to both a 12.5 year PTA observation and LISA for 100 realizations of a simulated SMBHB population for models A-D. Given the low rate of strong burst strains originating from PTA target populations, it is unsurprising that memory events with $\rm SNR\geq5$ occur $\num{9.8}{-8} - \num{1.2}{-6}$ times per century in the optimistic model, and less than once per Hubble time in the most pessimistic model. LISA demonstrates better prospects with $\rm SNR\geq5$ events occurring $0.3 - 2.8$ times per year in the most optimistic model, and less than once per million years in the most pessimistic.

\begin{figure}[htb]
\centering

        {
            \includegraphics[width=\columnwidth]{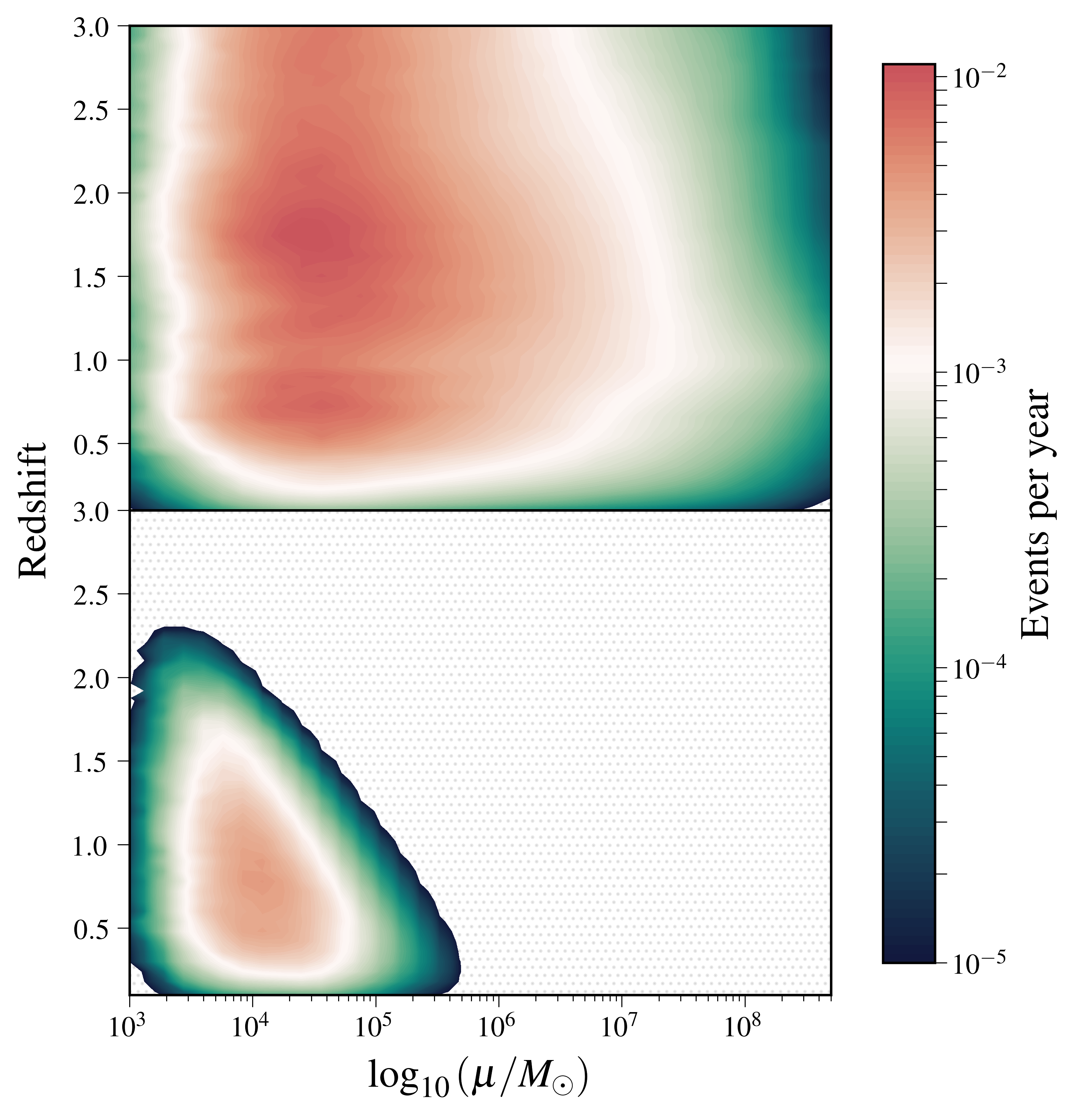}
        }    

    \caption{Mean memory burst rate by burst redshift and MBBH reduced mass. Upper panel is model A with $\alpha=3.0$; bottom panel is model C. Total number of memory events 3.3 and 0.4 per year, respectively.}
\label{fig:rate_contour}
\end{figure}

We break down the event rates for this simulation by reduced mass and burst redshift in Figure~\ref{fig:rate_contour}. We find the most frequent binary coalescences to occur among reduced masses between $10^{3}\,\msun$ and $10^{6}\,\msun$, constituting 99\% of all memory events. Such a region of binary mass is relatively unprobed. The last few years have seen numerous searches for active galactic nuclei from the black holes in these local, low-mass galaxies and future time-domain surveys, like the Large Synoptic Survey Telescope, may be able to constrain this population \citep{Greene+2018}. Memory detections from these sources would therefore complement our understanding of LISA's target population. As well, studies indicate discernible electromagnetic (EM) counterparts from these coalescences \citep{Tang+2018}. Paired with an EM trigger, memory signal searches could be conducted with restricted parameter priors to boost detection SNRs.

As we compare models A and C (upper and lower panel of Fig.~\ref{fig:rate_contour}, respectively), we see that decreasing environmental efficiency results in higher-mass binaries stalled at significant rates. The total number of bursts across our parameter space drops from 3.3 to 0.4 times per year. Lower-mass binaries initiated near the limits of parameter space evolve to closer redshifts, making up for those which may have stalled and keeping $0.1 < z < 1.5$ consistently populated. As shown in \cite{Sesana+2007}, most MBHBs coalescences whose inspiral signature yields LISA SNR > 5 lie beyond $z=3$ for a 3-year LISA lifetime, meaning the observation-based results reported here may be interpreted as lower limits on the number LISA memory events. 

\section{Outlook and Conclusions}
\label{sec:conclusions}
We evaluated the prospects for detecting a GW memory burst from source populations within reach of low-frequency GW detectors. We modified the initial inputs from population synthesis to include galaxy masses below $10^{10} \,\msun$. This probes additional SMBHB masses down to $10^{4}\,\msun$, and includes a region of observable orphan coalescences from parent mergers outside the LISA band. 
The incorporation of the decoupling radius allowed us to take into account the efficiency of the environmentally-driven orbital decay. Even assuming highly efficient environments, PTAs suffer from a sparser population of coalescing SMBHBs, making memory detection from these sources unlikely. Projections of PTA sensitivity at the advent of the LISA era strongly depend on the likelihood of PTA's access to increasingly sensitive radio telescopes, such as the planned ngVLA, SKA and DSA2000 \citep{ngVLA2018, Wang2018, DSA-2000}. Detailed explorations of future PTA sensitivities are beyond the scope of this work, although SNRs $\geq 5$ remain infrequent at $\num{1.7}{-3}$ per century for even a $30$ year PTA data set. 

SNRs $\geq 5$ can occur on average more than $0.8\,\rm yr^{-1}$ for LISA. Considering an $M_{\rm BH}-\mbulge$ selection-bias, rates diminish, although it is likely that expanding our parameter space to include more lower-mass SMBHBs beyond $z=3$ will boost our estimates. Here, we restricted our restricted ourselves to within currently resolvable astrophysical parameter values. In this context, if $< 10^{6}\,\msun$ binaries are numerous and indeed driven to low separations before their orbital decay is dominated by GW emission, LISA has a compelling science case in GW memory.

\textit{Acknowledgments}. --
We'd like to extend thanks to those who have helped in the production of this paper. Particularly Marc Favata, Michele Vallisneri, David Kaplan, Luke Zoltan Kelley, Dusty Madison, and Sarah Vigeland for helpful communications and discussions. We are grateful for computational resources provided by the Leonard E Parker Center for Gravitation, Cosmology and Astrophysics at the University of Wisconsin-Milwaukee. Part of this research was carried out at the Jet Propulsion Laboratory, California Institute of Technology, under a contract with the National Aeronautics and Space Administration.
The NANOGrav project receives support from National Science Foundation (NSF) Physics Frontiers Center award number 1430284.
\newpage
\bibliographystyle{yahapj}
\bibliography{apjjabb,biblio}

\end{document}